\documentclass[11pt]{article}
\usepackage{graphicx}
\usepackage{amsmath}
\usepackage{amssymb}
\usepackage{caption2}
\begin{document}
\bibliographystyle{prsty}
\begin{center}
{\large {\bf \sc{  Analysis of  the $\Omega_c(3000)$, $\Omega_c(3050)$, $\Omega_c(3066)$, $\Omega_c(3090)$ and $\Omega_c(3119)$   with QCD sum rules }}} \\[2mm]
Zhi-Gang Wang \footnote{E-mail:zgwang@aliyun.com.  }     \\
 Department of Physics, North China Electric Power University,
Baoding 071003, P. R. China
\end{center}

\begin{abstract}
In this article, we assign the $\Omega_c(3000)$, $\Omega_c(3050)$, $\Omega_c(3066)$, $\Omega_c(3090)$ and $\Omega_c(3119)$  to be the  P-wave baryon states with $J^P={\frac{1}{2}}^-$, ${\frac{1}{2}}^-$, ${\frac{3}{2}}^-$, ${\frac{3}{2}}^-$ and ${\frac{5}{2}}^-$, respectively, and study them  with the QCD sum rules  by introducing an explicit relative P-wave between the two  $s$ quarks.  The predictions support assigning the $\Omega_c(3050)$, $\Omega_c(3066)$, $\Omega_c(3090)$ and $\Omega_c(3119)$  to be the  P-wave baryon states with $J^P={\frac{1}{2}}^-$, ${\frac{3}{2}}^-$, ${\frac{3}{2}}^-$ and ${\frac{5}{2}}^-$, respectively, where the two $s$ quarks are in relative P-wave; while the  $\Omega_c(3000)$ can be assigned to the P-wave baryon state with $J^{P}={\frac{1}{2}}^-$, where the two $s$ quarks are in relative S-wave.
\end{abstract}



\section{Introduction}

In the past years,   several new charmed baryon states have been observed,  and    the
  spectroscopy of the charmed  baryon states have re-attracted  much attentions \cite{PDG}, the QCD sum rules plays  an important roles in assigning those new baryon states.  The masses of the  heavy baryon states with $J^P={\frac{1}{2}}^\pm$, ${\frac{3}{2}}^\pm$, ${\frac{5}{2}}^\pm$
 have been studied with the full QCD sum rules \cite{HeavyB-QCDSR,WangHbaryon, WangNegativeP,WangNegativeCTP} or
 the QCD sum rules combined with the heavy quark effective theory  \cite{HBE-QCDSR}.

Recently, the LHCb collaboration studied the  $\Xi_c^+ K^-$  mass spectrum  with a sample of $pp$  collision data corresponding to an integrated luminosity of  $3.3\rm{fb}^{-1}$   collected by the LHCb experiment, and observed five new narrow excited $\Omega_c^0$ states,
$\Omega_c(3000)$, $\Omega_c(3050)$, $\Omega_c(3066)$, $\Omega_c(3090)$, $\Omega_c(3119)$ \cite{LHCb-Omega}. The measured masses and widths are
\begin{flalign}
 & \Omega_c(3000) : M = 3000.4 \pm 0.2 \pm 0.1 \mbox{ MeV}\, , \, \Gamma = 4.5\pm0.6\pm0.3 \mbox{ MeV} \, , \nonumber \\
 & \Omega_c(3050) : M = 3050.2 \pm 0.1 \pm 0.1 \mbox{ MeV}\, , \, \Gamma = 0.8\pm0.2\pm0.1 \mbox{ MeV} \, , \nonumber \\
 & \Omega_c(3066) : M = 3065.6 \pm 0.1 \pm 0.3 \mbox{ MeV}\, , \, \Gamma = 3.5\pm0.4\pm0.2 \mbox{ MeV} \, , \nonumber \\
 & \Omega_c(3090) : M = 3090.2 \pm 0.3 \pm 0.5 \mbox{ MeV}\, , \, \Gamma = 8.7\pm1.0\pm0.8 \mbox{ MeV} \, , \nonumber \\
 & \Omega_c(3119) : M = 3119.1 \pm 0.3 \pm 0.9 \mbox{ MeV}\, , \, \Gamma = 1.1\pm0.8\pm0.4 \mbox{ MeV} \, .
\end{flalign}
There have been several assignments for those new charmed states. In Ref.\cite{Azizi-Omega}, the $\Omega_c(3066)$ and $\Omega_c(3119)$ are assigned to be the
2S $\Omega_c^0$  states  with $J^P={\frac{1}{2}}^+$ and ${\frac{3}{2}}^+$, respectively.
In Ref.\cite{Chen-Omega},  possible assignments of those $\Omega_c^0$ states to be the P-wave baryon states with $J^P={\frac{1}{2}}^-$, ${\frac{3}{2}}^-$ and ${\frac{5}{2}}^-$ are discussed.
In Refs.\cite{Rosner-Omega,WangZhu-Omega,Mathur-Omega}, the $\Omega_c(3000)$, $\Omega_c(3050)$, $\Omega_c(3066)$, $\Omega_c(3090)$ and $\Omega_c(3119)$ are assigned to be the  P-wave baryon states with $J^P={\frac{1}{2}}^-$, ${\frac{1}{2}}^-$, ${\frac{3}{2}}^-$, ${\frac{3}{2}}^-$ and ${\frac{5}{2}}^-$, respectively.
In Ref.\cite{Ping-Omega}, those $\Omega_c^0$ states are assigned to be the pentaquark states or molecular pentaquark states with $J^P={\frac{1}{2}}^-$, ${\frac{3}{2}}^-$ or ${\frac{5}{2}}^-$.
In Ref.\cite{Zhong-Omega}, the $\Omega_c(3000)$, $\Omega_c(3050)$, $\Omega_c(3066)$ and $\Omega_c(3090)$  are assigned to be the  P-wave baryon states with $J^P={\frac{1}{2}}^-$,  ${\frac{3}{2}}^-$, ${\frac{3}{2}}^-$ and ${\frac{1}{2}}^-$, respectively.
In Ref.\cite{Cheng-Omega}, the $\Omega_c(3090)$ and $\Omega_c(3119)$ are assigned to be the
2S  $\Omega_c^0$ states with $J^P={\frac{1}{2}}^+$ and ${\frac{3}{2}}^+$, respectively, while the $\Omega_c(3000)$, $\Omega_c(3066)$ and $\Omega_c(3050)$ are assigned to be the P-wave baryon states with $J^P={\frac{1}{2}}^-$,  ${\frac{3}{2}}^-$ and ${\frac{5}{2}}^-$, respectively.

In this article, we tentatively assign the $\Omega_c(3000)$, $\Omega_c(3050)$, $\Omega_c(3066)$, $\Omega_c(3090)$ and $\Omega_c(3119)$  to be the  P-wave baryon states with $J^P={\frac{1}{2}}^-$, ${\frac{1}{2}}^-$, ${\frac{3}{2}}^-$, ${\frac{3}{2}}^-$ and ${\frac{5}{2}}^-$, respectively, and study their  masses and pole residues with the QCD sum rules in details.

 The ground state quarks have the spin-parity ${1\over 2}^+$, two
quarks can form a scalar diquark or an axialvector diquark with the
spin-parity $0^+$ or $1^+$, the
diquark then combines  with a third quark to form a positive parity
baryon with spin ${1\over 2}$ or ${3\over 2}$. We can construct the baryon currents $\eta$ and $\eta_\mu$ with positive parity without introducing additional P-wave.
 As multiplying $i \gamma_{5}$ to the baryon currents  changes their parity, the currents $i \gamma_{5}\eta$  and $i \gamma_{5}\eta_\mu$ couple potentially to the negative parity heavy baryon states.
In Refs.\cite{WangNegativeP,WangNegativeCTP}, we construct the currents without introducing relative P-wave to study the negative parity heavy, doubly-heavy and triply-heavy   baryon states, and obtain satisfactory results. The predictions $M=2.98\pm0.16 \,\rm{GeV}$ for the $\Omega_c^0$ states with $J^P={\frac{1}{2}}^-$, ${\frac{3}{2}}^-$ are consistent with the masses of the  $\Omega_c(3000)$, $\Omega_c(3050)$, $\Omega_c(3066)$, $\Omega_c(3090)$ from the LHCb collaboration \cite{WangNegativeP}.

 In Ref.\cite{Wang-2625-2815}, we construct the interpolating currents by introducing the relative P-wave explicitly,  study the  negative parity charmed baryon states $\Lambda_c(2625)$ and $\Xi_c(2815)$ with  the full QCD sum rules, and reproduce the experimental values of the masses. In this article, we extend our previous work to study the $\Omega_c(3000)$, $\Omega_c(3050)$, $\Omega_c(3066)$, $\Omega_c(3090)$ and $\Omega_c(3119)$ with QCD sum rules by introducing the relative P-wave explicitly.

 The article is arranged as follows:  we derive the QCD sum rules for the masses and pole residues of  the $\Omega_c^0$ states in Sect.2;
 in Sect.3, we present the numerical results and discussions; and Sect.4 is reserved for our
conclusions.

\section{QCD sum rules for  the $\Omega_c^0$ states}

In the following, we write down  the two-point correlation functions $\Pi(p)$, $\Pi_{\mu\nu}(p)$, $\Pi_{\mu\nu\alpha\beta}(p)$  in the QCD sum rules,
\begin{eqnarray}
\Pi(p)&=&i\int d^4x e^{ip \cdot x} \langle0|T\left\{J(x)\bar{J}(0)\right\}|0\rangle \, , \nonumber\\
\Pi_{\mu\nu}(p)&=&i\int d^4x e^{ip \cdot x} \langle0|T\left\{J_{\mu}(x)\bar{J}_{\nu}(0)\right\}|0\rangle \, , \nonumber\\
\Pi_{\mu\nu\alpha\beta}(p)&=&i\int d^4x e^{ip \cdot x} \langle0|T\left\{J_{\mu\nu}(x)\bar{J}_{\alpha\beta}(0)\right\}|0\rangle \, ,
\end{eqnarray}
where $J(x)=J^1(x),\,J^2(x)$, $J_\mu(x)=J^1_\mu(x),\,J^2_\mu(x)$,
\begin{eqnarray}
J^1(x)&=&i\varepsilon^{ijk} \left[ \partial^\mu s^T_i(x) C\gamma^\nu s_j(x)+ s^T_i(x) C\gamma^\nu \partial^{\mu}s_j(x)\right]g_{\mu\nu}\,c_k(x) \, , \nonumber\\
J^2(x)&=&i\varepsilon^{ijk} \left[ \partial^\mu s^T_i(x) C\gamma^\nu s_j(x)+ s^T_i(x) C\gamma^\nu \partial^{\mu}s_j(x)\right]\sigma_{\mu\nu}\,c_k(x) \, ,\nonumber \\
J^1_{\mu}(x)&=&i\varepsilon^{ijk} \left[ \partial^\alpha s^T_i(x) C\gamma^\beta s_j(x)+s^T_i(x) C\gamma^\beta \partial^{\alpha}s_j(x)\right]\left(\widetilde{g}_{\mu\alpha}\gamma_\beta-\widetilde{g}_{\mu\beta}\gamma_\alpha \right)\gamma_5 c_k(x) \, ,\nonumber \\
J^2_{\mu}(x)&=&i\varepsilon^{ijk} \left[ \partial^\alpha s^T_i(x) C\gamma^\beta s_j(x)+ s^T_i(x) C\gamma^\beta \partial^{\alpha}s_j(x)\right]\nonumber\\
&&\left(g_{\mu\alpha}\gamma_\beta+g_{\mu\beta}\gamma_\alpha-\frac{1}{2}g_{\alpha\beta}\gamma_\mu \right)\gamma_5 c_k(x) \, ,\nonumber\\
J_{\mu\nu}(x)&=&i\varepsilon^{ijk} \left[ \partial_\mu s^T_i(x) C\gamma_\nu s_j(x)+\partial_\nu s^T_i(x) C\gamma_\mu s_j(x)+ s^T_i(x) C\gamma_\nu \partial_{\mu}s_j(x)\right.\nonumber\\
&&\left.+ s^T_i(x) C\gamma_\mu \partial_{\nu}s_j(x)\right] c_k(x) \, ,
\end{eqnarray}
$\widetilde{g}_{\mu\nu}=g_{\mu\nu}-\frac{1}{4}\gamma_\mu\gamma_\nu$,
the $i$, $j$, $k$ are color indices, the $C$ is the charge conjugation matrix.
We construct the  currents with the light diquarks $S^{i}_{\mu\nu}=\varepsilon^{ijk} \left[ \partial_\mu s^T_i C\gamma_\nu s_j+ s^T_i C\gamma_\nu \partial_{\mu}s_j\right]$.
The $S^{i}_{\mu\nu}$
  have two Lorentz indices $\mu$ and $\nu$, but they  are neither symmetric nor   anti-symmetric when interchanging
  the indices $\mu$ and $\nu$.  The scalar components $S^{i}_{\mu\nu}g^{\mu\nu}$ and $S^{i}_{\mu\nu}\sigma^{\mu\nu}$  couple potentially to the spin-0 diquarks.
  The Dirac matrixes $\widetilde{g}^{\alpha\mu}\gamma^\nu-\widetilde{g}^{\alpha\nu}\gamma^\mu $ and $g^{\alpha\mu}\gamma^\nu+g^{\alpha\nu}\gamma^\mu-\frac{1}{2}g^{\mu\nu}\gamma^\alpha$ are anti-symmetric and symmetric respectively when interchanging
  the indices $\mu$ and $\nu$, the vector  components $S^{i}_{\mu\nu}\left(\widetilde{g}^{\alpha\mu}\gamma^\nu-\widetilde{g}^{\alpha\nu}\gamma^\mu\right) $ and $S^{i}_{\mu\nu}\left(g^{\alpha\mu}\gamma^\nu+g^{\alpha\nu}\gamma^\mu-\frac{1}{2}g^{\mu\nu}\gamma^\alpha\right)$ couple potentially to the spin-1 diquarks. The symmetric components $S^{i}_{\mu\nu}+S^{i}_{\nu\mu}$ couple potentially to the spin-0 and 2 diquarks. So we choose the currents $J(x)$, $J_\mu(x)$ and $J_{\mu\nu}(x)$ to study the spin-$\frac{1}{2}$, $\frac{3}{2}$ and $\frac{5}{2}$ baryon states, respectively.

 The currents $J(0)$, $J_\mu(0)$ and $J_{\mu\nu}(0)$ couple potentially to the ${\frac{1}{2}}^-$, ${\frac{1}{2}}^+$, ${\frac{3}{2}}^-$ and ${\frac{1}{2}}^-$, ${\frac{3}{2}}^+$, ${\frac{5}{2}}^-$  charmed baryon
 states $B_{\frac{1}{2}}^{-}$, $B_{\frac{1}{2}}^{+}$, $B_{\frac{3}{2}}^{-}$ and $B_{\frac{1}{2}}^{-}$, $B_{\frac{3}{2}}^{+}$, $B_{\frac{5}{2}}^{-}$, respectively,
\begin{eqnarray}
\langle 0| J (0)|B_{\frac{1}{2}}^{-}(p)\rangle &=&\lambda^{-}_{\frac{1}{2}} U^{-}(p,s) \, ,  \\
\langle 0| J_{\mu} (0)|B_{\frac{1}{2}}^{+}(p)\rangle &=&f^{+}_{\frac{1}{2}}p_\mu U^{+}(p,s) \, , \nonumber \\
\langle 0| J_{\mu} (0)|B_{\frac{3}{2}}^{-}(p)\rangle &=&\lambda^{-}_{\frac{3}{2}} U^{-}_\mu(p,s) \, ,  \\
\langle 0| J_{\mu\nu} (0)|B_{\frac{1}{2}}^{-}(p)\rangle &=&g^{-}_{\frac{1}{2}}p_\mu p_\nu U^{-}(p,s) \, , \nonumber\\
\langle 0| J_{\mu\nu} (0)|B_{\frac{3}{2}}^{+}(p)\rangle &=&f^{+}_{\frac{3}{2}} \left[p_\mu U^{+}_{\nu}(p,s)+p_\nu U^{+}_{\mu}(p,s)\right] \, , \nonumber\\
\langle 0| J_{\mu\nu} (0)|B_{\frac{5}{2}}^{-}(p)\rangle &=&\lambda^{-}_{\frac{5}{2}} U^{-}_{\mu\nu}(p,s) \, .
\end{eqnarray}
On the other hand, the currents  $J(0)$, $J_\mu(0)$ and $J_{\mu\nu}(0)$ couple potentially to the ${\frac{1}{2}}^+$, ${\frac{1}{2}}^-$, ${\frac{3}{2}}^+$ and ${\frac{1}{2}}^+$, ${\frac{3}{2}}^-$, ${\frac{5}{2}}^+$  charmed baryon
 states $B_{\frac{1}{2}}^{+}$, $B_{\frac{1}{2}}^{-}$, $B_{\frac{3}{2}}^{+}$ and $B_{\frac{1}{2}}^{+}$, $B_{\frac{3}{2}}^{-}$, $B_{\frac{5}{2}}^{+}$, respectively \cite{Oka96,WangPc},
 \begin{eqnarray}
 \langle 0| J (0)|B_{\frac{1}{2}}^{+}(p)\rangle &=&\lambda^{+}_{\frac{1}{2}}  i\gamma_5 U^{+}(p,s) \, ,  \\
 \langle 0| J_{\mu} (0)|B_{\frac{1}{2}}^{-}(p)\rangle &=&f^{-}_{\frac{1}{2}}p_\mu i\gamma_5 U^{-}(p,s) \, , \nonumber\\
\langle 0| J_{\mu} (0)|B_{\frac{3}{2}}^{+}(p)\rangle &=&\lambda^{+}_{\frac{3}{2}}i\gamma_5 U^{+}_{\mu}(p,s) \, , \\
\langle 0| J_{\mu\nu} (0)|B_{\frac{1}{2}}^{+}(p)\rangle &=&g^{+}_{\frac{1}{2}}p_\mu p_\nu i\gamma_5 U^{+}(p,s) \, , \nonumber\\
\langle 0| J_{\mu\nu} (0)|B_{\frac{3}{2}}^{-}(p)\rangle &=&f^{-}_{\frac{3}{2}} i\gamma_5\left[p_\mu U^{-}_{\nu}(p,s)+p_\nu U^{-}_{\mu}(p,s)\right] \, , \nonumber\\
\langle 0| J_{\mu\nu} (0)|B_{\frac{5}{2}}^{+}(p)\rangle &=&\lambda^{+}_{\frac{5}{2}}i\gamma_5 U^{+}_{\mu\nu}(p,s) \, .
\end{eqnarray}
The spinors $U^\pm(p,s)$ satisfy the Dirac equations  $(\not\!\!p-M_{\pm})U^{\pm}(p)=0$, while the spinors $U^{\pm}_\mu(p,s)$ and $U^{\pm}_{\mu\nu}(p,s)$ satisfy the Rarita-Schwinger equations $(\not\!\!p-M_{\pm})U^{\pm}_\mu(p)=0$ and $(\not\!\!p-M_{\pm})U^{\pm}_{\mu\nu}(p)=0$,  and the relations $\gamma^\mu U^{\pm}_\mu(p,s)=0$,
$p^\mu U^{\pm}_\mu(p,s)=0$, $\gamma^\mu U^{\pm}_{\mu\nu}(p,s)=0$,
$p^\mu U^{\pm}_{\mu\nu}(p,s)=0$, $ U^{\pm}_{\mu\nu}(p,s)= U^{\pm}_{\nu\mu}(p,s)$. The $\lambda^{\pm}_{\frac{1}{2}/\frac{3}{2}/\frac{5}{2}}$, $f^{\pm}_{\frac{1}{2}/\frac{3}{2}}$ and $g^{\pm}_{\frac{1}{2}}$ are the pole residues or current-baryon coupling constants.

At the phenomenological side,  we  insert  a complete set  of intermediate charmed baryon  states with the
same quantum numbers as the current operators $J(x)$,
$i\gamma_5 J(x)$, $J_\mu(x)$,
$i\gamma_5 J_\mu(x)$, $J_{\mu\nu}(x)$ and
$i\gamma_5 J_{\mu\nu}(x)$ into the correlation functions $\Pi(p)$,
$\Pi_{\mu\nu}(p)$ and $\Pi_{\mu\nu\alpha\beta}(p)$ to obtain the hadronic representation
\cite{SVZ79,PRT85}. After isolating the pole terms of the lowest
states of the charmed  baryon states, we obtain the
following results:
\begin{eqnarray}
 \Pi(p) & = & {\lambda^{-}_{\frac{1}{2}}}^2  {\!\not\!{p}+ M_{-} \over M_{-}^{2}-p^{2}  }
 +  {\lambda^{+}_{\frac{1}{2}}}^2  {\!\not\!{p}- M_{+} \over M_{+}^{2}-p^{2}  } +\cdots  \, , \\
  \Pi_{\mu\nu}(p) & = & {\lambda^{-}_{\frac{3}{2}}}^2  {\!\not\!{p}+ M_{-} \over M_{-}^{2}-p^{2}  } \left(- g_{\mu\nu}+\frac{\gamma_\mu\gamma_\nu}{3}+\frac{2p_\mu p_\nu}{3p^2}-\frac{p_\mu\gamma_\nu-p_\nu \gamma_\mu}{3\sqrt{p^2}}
\right)\nonumber\\
&&+  {\lambda^{+}_{\frac{3}{2}}}^2  {\!\not\!{p}- M_{+} \over M_{+}^{2}-p^{2}  } \left(- g_{\mu\nu}+\frac{\gamma_\mu\gamma_\nu}{3}+\frac{2p_\mu p_\nu}{3p^2}-\frac{p_\mu\gamma_\nu-p_\nu \gamma_\mu}{3\sqrt{p^2}}
\right)   \nonumber \\
& &+ {f^{+}_{\frac{1}{2}}}^2  {\!\not\!{p}+ M_{+} \over M_{+}^{2}-p^{2}  } p_\mu p_\nu+  {f^{-}_{\frac{1}{2}}}^2  {\!\not\!{p}- M_{-} \over M_{-}^{2}-p^{2}  } p_\mu p_\nu  +\cdots  \, ,
\end{eqnarray}
\begin{eqnarray}
\Pi_{\mu\nu\alpha\beta}(p) & = & {\lambda^{-}_{\frac{5}{2}}}^2  {\!\not\!{p}+ M_{-} \over M_{-}^{2}-p^{2}  } \left[\frac{ \widetilde{g}_{\mu\alpha}\widetilde{g}_{\nu\beta}+\widetilde{g}_{\mu\beta}\widetilde{g}_{\nu\alpha}}{2}-\frac{\widetilde{g}_{\mu\nu}\widetilde{g}_{\alpha\beta}}{5}-\frac{1}{10}\left( \gamma_{\mu}\gamma_{\alpha}+\frac{\gamma_{\mu}p_{\alpha}-\gamma_{\alpha}p_{\mu}}{\sqrt{p^2}}-\frac{p_{\mu}p_{\alpha}}{p^2}\right)\widetilde{g}_{\nu\beta}\right.\nonumber\\
&&\left.-\frac{1}{10}\left( \gamma_{\nu}\gamma_{\alpha}+\frac{\gamma_{\nu}p_{\alpha}-\gamma_{\alpha}p_{\nu}}{\sqrt{p^2}}-\frac{p_{\nu}p_{\alpha}}{p^2}\right)\widetilde{g}_{\mu\beta}
+\cdots\right]\nonumber\\
&&+   {\lambda^{+}_{\frac{5}{2}}}^2  {\!\not\!{p}- M_{+} \over M_{+}^{2}-p^{2}  } \left[\frac{ \widetilde{g}_{\mu\alpha}\widetilde{g}_{\nu\beta}+\widetilde{g}_{\mu\beta}\widetilde{g}_{\nu\alpha}}{2}
-\frac{\widetilde{g}_{\mu\nu}\widetilde{g}_{\alpha\beta}}{5}-\frac{1}{10}\left( \gamma_{\mu}\gamma_{\alpha}+\frac{\gamma_{\mu}p_{\alpha}-\gamma_{\alpha}p_{\mu}}{\sqrt{p^2}}-\frac{p_{\mu}p_{\alpha}}{p^2}\right)\widetilde{g}_{\nu\beta}\right.\nonumber\\
&&\left.
-\frac{1}{10}\left( \gamma_{\nu}\gamma_{\alpha}+\frac{\gamma_{\nu}p_{\alpha}-\gamma_{\alpha}p_{\nu}}{\sqrt{p^2}}-\frac{p_{\nu}p_{\alpha}}{p^2}\right)\widetilde{g}_{\mu\beta}
 +\cdots\right]   \nonumber\\
 && +{f^{+}_{\frac{3}{2}}}^2  {\!\not\!{p}+ M_{+} \over M_{+}^{2}-p^{2}  } \left[ p_\mu p_\alpha \left(- g_{\nu\beta}+\frac{\gamma_\nu\gamma_\beta}{3}+\frac{2p_\nu p_\beta}{3p^2}-\frac{p_\nu\gamma_\beta-p_\beta \gamma_\nu}{3\sqrt{p^2}}
\right)+\cdots \right]\nonumber\\
&&+  {f^{-}_{\frac{3}{2}}}^2  {\!\not\!{p}- M_{-} \over M_{-}^{2}-p^{2}  } \left[ p_\mu p_\alpha \left(- g_{\nu\beta}+\frac{\gamma_\nu\gamma_\beta}{3}+\frac{2p_\nu p_\beta}{3p^2}-\frac{p_\nu\gamma_\beta-p_\beta \gamma_\nu}{3\sqrt{p^2}}
\right)+\cdots \right]   \nonumber \\
& &+ {g^{-}_{\frac{1}{2}}}^2  {\!\not\!{p}+ M_{-} \over M_{-}^{2}-p^{2}  } p_\mu p_\nu p_\alpha p_\beta+  {g^{+}_{\frac{1}{2}}}^2  {\!\not\!{p}- M_{+} \over M_{+}^{2}-p^{2}  } p_\mu p_\nu p_\alpha p_\beta  +\cdots \, ,
    \end{eqnarray}
where $\widetilde{g}_{\mu\nu}=g_{\mu\nu}-\frac{p_{\mu}p_{\nu}}{p^2}$.
In calculations, we have used the following summations \cite{HuangShiZhong},
\begin{eqnarray}
\sum_s U \overline{U}&=&\left(\!\not\!{p}+M_{\pm}\right) \,  ,  \\
\sum_s U_\mu \overline{U}_\nu&=&\left(\!\not\!{p}+M_{\pm}\right)\left( -g_{\mu\nu}+\frac{\gamma_\mu\gamma_\nu}{3}+\frac{2p_\mu p_\nu}{3p^2}-\frac{p_\mu
\gamma_\nu-p_\nu \gamma_\mu}{3\sqrt{p^2}} \right) \,  ,  \\
\sum_s U_{\mu\nu}\overline {U}_{\alpha\beta}&=&\left(\!\not\!{p}+M_{\pm}\right)\left\{\frac{\widetilde{g}_{\mu\alpha}\widetilde{g}_{\nu\beta}+\widetilde{g}_{\mu\beta}\widetilde{g}_{\nu\alpha}}{2} -\frac{\widetilde{g}_{\mu\nu}\widetilde{g}_{\alpha\beta}}{5}-\frac{1}{10}\left( \gamma_{\mu}\gamma_{\alpha}+\frac{\gamma_{\mu}p_{\alpha}-\gamma_{\alpha}p_{\mu}}{\sqrt{p^2}}-\frac{p_{\mu}p_{\alpha}}{p^2}\right)\widetilde{g}_{\nu\beta}\right. \nonumber\\
&&-\frac{1}{10}\left( \gamma_{\nu}\gamma_{\alpha}+\frac{\gamma_{\nu}p_{\alpha}-\gamma_{\alpha}p_{\nu}}{\sqrt{p^2}}-\frac{p_{\nu}p_{\alpha}}{p^2}\right)\widetilde{g}_{\mu\beta}
-\frac{1}{10}\left( \gamma_{\mu}\gamma_{\beta}+\frac{\gamma_{\mu}p_{\beta}-\gamma_{\beta}p_{\mu}}{\sqrt{p^2}}-\frac{p_{\mu}p_{\beta}}{p^2}\right)\widetilde{g}_{\nu\alpha}\nonumber\\
&&\left.-\frac{1}{10}\left( \gamma_{\nu}\gamma_{\beta}+\frac{\gamma_{\nu}p_{\beta}-\gamma_{\beta}p_{\nu}}{\sqrt{p^2}}-\frac{p_{\nu}p_{\beta}}{p^2}\right)\widetilde{g}_{\mu\alpha} \right\} \, ,
\end{eqnarray}
and $p^2=M^2_{\pm}$ on the mass-shell.

We can rewrite the correlation functions $\Pi(p)$, $\Pi_{\mu\nu}(p)$ and $\Pi_{\mu\nu\alpha\beta}(p)$ into the following form according to Lorentz covariance,
\begin{eqnarray}
\Pi(p)&=&\Pi_{\frac{1}{2}}(p^2)\, , \\
\Pi_{\mu\nu}(p)&=&\Pi_{\frac{3}{2}}(p^2)\,\left(- g_{\mu\nu}\right)+\cdots\, , \\
\Pi_{\mu\nu\alpha\beta}(p)&=&\Pi_{\frac{5}{2}}(p^2)\,\frac{ g_{\mu\alpha}g_{\nu\beta}+g_{\mu\beta}g_{\nu\alpha}}{2}+\cdots \, .
\end{eqnarray}
 In this article, we choose the tensor structures $g_{\mu\nu}$ and $g_{\mu\alpha}g_{\nu\beta}+g_{\mu\beta}g_{\nu\alpha}$ for analysis, and separate the contributions of the ${\frac{3}{2}}^{\pm}$ and ${\frac{5}{2}}^{\pm}$ charmed baryon states unambiguously. For detailed discussions on this subject, one can consult Ref.\cite{WangPc}.

We obtain the hadronic spectral densities at phenomenological side through the dispersion relation,
\begin{eqnarray}
\frac{{\rm Im}\Pi_{j}(s)}{\pi}&=&\!\not\!{p} \left[{\lambda^{-}_{j}}^2 \delta\left(s-M_{-}^2\right)+{\lambda^{+}_{j}}^2 \delta\left(s-M_{+}^2\right)\right] +\left[M_{-}{\lambda^{-}_{j}}^2 \delta\left(s-M_{-}^2\right)-M_{+}{\lambda^{+}_{j}}^2 \delta\left(s-M_{+}^2\right)\right]\, , \nonumber\\
&=&\!\not\!{p}\, \rho^1_{j,H}(s)+\rho^0_{j,H}(s) \, ,
\end{eqnarray}
where $j=\frac{1}{2}$, $\frac{3}{2}$, $\frac{5}{2}$, the subscript $H$ denotes  the hadron side,
then we introduce the weight function $\exp\left(-\frac{s}{T^2}\right)$ to obtain the QCD sum rules at the phenomenological side,
\begin{eqnarray}
\int_{m_c^2}^{s_0}ds \left[\sqrt{s}\rho^1_{j,H}(s)+\rho^0_{j,H}(s)\right]\exp\left( -\frac{s}{T^2}\right)
&=&2M_{-}{\lambda^{-}_{j}}^2\exp\left( -\frac{M_{-}^2}{T^2}\right) \, ,
\end{eqnarray}
where the $s_0$ are the continuum thresholds and the $T^2$ are the Borel parameters \cite{WangPc}.

At the QCD side, we  calculate the light quark parts of the correlation functions $\Pi(p)$,
 $\Pi_{\mu\nu}(p)$, $\Pi_{\mu\nu\alpha\beta}(p)$ with the full light quark propagators   in the coordinate space and
take  the momentum space expression for the full $c$-quark propagator. It is straightforward but tedious to compute  the integrals both in the coordinate and momentum spaces  to obtain the correlation functions $\Pi_{j}(p^2)$, therefore the QCD spectral densities  through  the dispersion relation,
\begin{eqnarray}
\frac{{\rm Im}\Pi_{j}(s)}{\pi}&=&\!\not\!{p}\, \rho^1_{j,QCD}(s)+\rho^0_{j,QCD}(s) \, ,
\end{eqnarray}
where $j=\frac{1}{2}$, $\frac{3}{2}$, $\frac{5}{2}$, the explicit expressions of the QCD spectral densities $\rho^1_{j,QCD}(s)$ and $\rho^0_{j,QCD}(s)$ are neglected for simplicity. In this article, we carry out the operator product expansion up to the vacuum condensates of dimension 10 and take into account  the vacuum condensates $\langle\bar{s}s\rangle$, $\langle \frac{\alpha_sGG}{\pi}\rangle$, $\langle\bar{s}g_s\sigma Gs\rangle$, $\langle\bar{s}s\rangle\langle\bar{s}g_s\sigma Gs\rangle$, $\langle\bar{s}g_s\sigma Gs\rangle^2$.

Once the analytical QCD spectral densities $\rho^1_{j,QCD}(s)$ and $\rho^0_{j,QCD}(s)$ are obtained,  we can take the
quark-hadron duality below the continuum thresholds  $s_0$ and introduce the weight function $\exp\left(-\frac{s}{T^2}\right)$ to obtain  the  QCD sum rules:
\begin{eqnarray}
2M_{-}{\lambda^{-}_{j}}^2\exp\left( -\frac{M_{-}^2}{T^2}\right)
&=& \int_{m_c^2}^{s_0}ds \left[\sqrt{s}\rho^1_{j,QCD}(s)+\rho^0_{j,QCD}(s)\right]\exp\left( -\frac{s}{T^2}\right)\, .
\end{eqnarray}

We derive   Eq.(22) with respect to  $\frac{1}{T^2}$, then eliminate the
 pole residues $\lambda^{-}_j$ and obtain the QCD sum rules for
 the masses of the charmed baryon  states,
 \begin{eqnarray}
 M^2_{-} &=& \frac{-\frac{d}{d(1/T^2)}\int_{m_c^2}^{s_0}ds \left[\sqrt{s}\rho^1_{j,QCD}(s)+\rho^0_{j,QCD}(s)\right]\exp\left( -\frac{s}{T^2}\right)}{\int_{m_c^2}^{s_0}ds \left[\sqrt{s}\rho^1_{j,QCD}(s)+\rho^0_{j,QCD}(s)\right]\exp\left( -\frac{s}{T^2}\right)}\, .
\end{eqnarray}

\section{Numerical results and discussions}
The vacuum condensates are taken to be the standard values
$\langle\bar{q}q \rangle=-(0.24\pm 0.01\, \rm{GeV})^3$,  $\langle\bar{s}s \rangle=(0.8\pm0.1)\langle\bar{q}q \rangle$,
 $\langle\bar{s}g_s\sigma G s \rangle=m_0^2\langle \bar{s}s \rangle$,
$m_0^2=(0.8 \pm 0.1)\,\rm{GeV}^2$, $\langle \frac{\alpha_s
GG}{\pi}\rangle=(0.33\,\rm{GeV})^4 $    at the energy scale  $\mu=1\, \rm{GeV}$
\cite{SVZ79,PRT85,ColangeloReview}.
The quark condensate and mixed quark condensate evolve with the   renormalization group equation,
 $\langle\bar{s}s \rangle(\mu)=\langle\bar{s}s \rangle(Q)\left[\frac{\alpha_{s}(Q)}{\alpha_{s}(\mu)}\right]^{\frac{4}{9}}$,
 $\langle\bar{s}g_s \sigma Gs \rangle(\mu)=\langle\bar{s}g_s \sigma Gs \rangle(Q)\left[\frac{\alpha_{s}(Q)}{\alpha_{s}(\mu)}\right]^{\frac{2}{27}}$.

In the article, we take the $\overline{MS}$ masses $m_{c}(m_c)=(1.275\pm0.025)\,\rm{GeV}$ and $m_s(\mu=2\,\rm{GeV})=(0.095\pm0.005)\,\rm{GeV}$
 from the particle data group \cite{PDG}, and take into account
the energy-scale dependence of  the $\overline{MS}$ masses from the renormalization group equation,
\begin{eqnarray}
m_c(\mu)&=&m_c(m_c)\left[\frac{\alpha_{s}(\mu)}{\alpha_{s}(m_c)}\right]^{\frac{12}{25}} \, ,\nonumber\\
m_s(\mu)&=&m_s({\rm 2GeV} )\left[\frac{\alpha_{s}(\mu)}{\alpha_{s}({\rm 2GeV})}\right]^{\frac{4}{9}} \, ,\nonumber\\
\alpha_s(\mu)&=&\frac{1}{b_0t}\left[1-\frac{b_1}{b_0^2}\frac{\log t}{t} +\frac{b_1^2(\log^2{t}-\log{t}-1)+b_0b_2}{b_0^4t^2}\right]\, ,
\end{eqnarray}
  where $t=\log \frac{\mu^2}{\Lambda^2}$, $b_0=\frac{33-2n_f}{12\pi}$, $b_1=\frac{153-19n_f}{24\pi^2}$, $b_2=\frac{2857-\frac{5033}{9}n_f+\frac{325}{27}n_f^2}{128\pi^3}$,  $\Lambda=213\,\rm{MeV}$, $296\,\rm{MeV}$  and  $339\,\rm{MeV}$ for the flavors  $n_f=5$, $4$ and $3$, respectively  \cite{PDG}.

In Refs.\cite{WangTetraquark,WangMolecule}, we study the acceptable  energy scales of the QCD spectral densities  for the hidden-charm (bottom) tetraquark states and molecular  states in the QCD sum rules  for the first time,  and suggest an  empirical formula $\mu=\sqrt{M^2_{X/Y/Z}-(2{\mathbb{M}}_Q)^2}$ to determine  the optimal  energy scales, where the $X$, $Y$, $Z$ denote the four-quark states, and the ${\mathbb{M}}_Q$ is the effective heavy quark mass. The empirical energy scale formula also works well in studying the hidden-charm pentaquark states \cite{WangPc}.
  In Ref.\cite{Wang-2625-2815}, we use the diquark-quark model to construct the interpolating  currents, and take the analogous formula
$ \mu =\sqrt{M_{\Lambda_c/\Xi_c}^2-{\mathbb{M}}_c^2}$
   to determine the energy scales of the QCD spectral densities of the QCD sum rules for the charmed baryon states $\Lambda_c(2625)$ and $\Xi_c(2815)$, and obtain satisfactory results. In this article,  we use the formula $\mu=\sqrt{M_{\Omega_c}^2-{\mathbb{M}}_c^2}$ to determine the energy scales of the QCD spectral densities. If we take the updated value ${\mathbb{M}}_c=1.82\,\rm{GeV}$ \cite{WangEPJC4260}, then $\mu\approx2.5\,\rm{GeV}$. In calculations, we set the energy scales of the QCD spectral densities to be $\mu=2.5\,\rm{GeV}$.

Now  we search for the  Borel parameters $T^2$ and continuum threshold
parameters $s_0$  to satisfy the  following three  criteria:

$\bf{1_\cdot}$ Pole dominance at the phenomenological side;

$\bf{2_\cdot}$ Convergence of the operator product expansion;

$\bf{3_\cdot}$ Appearance of the Borel platforms.

In calculations, we observe that no stable QCD sum rules can be obtained for the current $J^2(x)$. The resulting   Borel parameters $T^2$, continuum threshold parameters $s_0$,  pole contributions   and  perturbative contributions (per)  are shown explicitly in Table 1, where the perturbative contributions  are defined by
 \begin{eqnarray}
 {\rm per} &=& \frac{\int_{m_c^2}^{s_0}ds\,\rho_{ per}(s)\exp\left( -\frac{s}{T^2}\right)}{\int_{m_c^2}^{s_0}ds\,\rho_{ tot}(s)\exp\left( -\frac{s}{T^2}\right)}\, \, ,
 \end{eqnarray}
  the $\rho_{ per}(s)$ and $\rho_{ tot}(s)$ denote the perturbative and total QCD spectral densities, respectively.
From the Table, we can see that the criteria $\bf 1$ and $\bf 2$ can be satisfied.

\begin{table}
\begin{center}
\begin{tabular}{|c|c|c|c|c|c|}\hline\hline
currents      & $T^2 (\rm{GeV}^2)$   & $\sqrt{s_0} (\rm{GeV})$    & pole          & perturbative \\  \hline
$J^1$         & $1.8-2.2$            & $3.6\pm0.1$                & $(43-73)\%$   & $(95-99)\%$ \\ \hline
$J^1_\mu$     & $1.8-2.2$            & $3.6\pm0.1$                & $(42-73)\%$   & $(100-104)\%$ \\ \hline
$J^2_\mu$     & $1.9-2.3$            & $3.6\pm0.1$                & $(43-72)\%$   & $(100-102)\%$ \\ \hline
$J_{\mu\nu}$  & $2.4-2.8$            & $3.7\pm0.1$                & $(42-66)\%$   & $(94-96)\%$ \\ \hline\hline
\end{tabular}
\end{center}
\caption{ The Borel parameters $T^2$, continuum threshold parameters $s_0$,
 pole contributions (pole)   and  perturbative contributions (perturbative).}
\end{table}

\begin{table}
\begin{center}
\begin{tabular}{|c|c|c|c|c|c|c|}\hline\hline
 currents       & $J_{j_l}^P$               & $M(\rm{GeV})$   &$\lambda (10^{-1}\rm{GeV}^4)$   & assignments \\ \hline
 $J^1$          & ${\frac{1}{2}}_{0}^-$     & $3.05\pm0.11$   &$2.34\pm0.50$                   & $\Omega_c(3050)$\\ \hline
 $J^1_\mu$      & ${\frac{3}{2}}_{1}^-$     & $3.06\pm0.11$   &$1.03\pm0.23$                   & $\Omega_c(3066/3090)$\\ \hline
 $J^2_\mu$      & ${\frac{3}{2}}_{2}^-$     & $3.06\pm0.10$   &$2.47\pm0.47$                   & $\Omega_c(3066/3090)$\\ \hline
 $J_{\mu\nu}$   & ${\frac{5}{2}}_{2}^-$     & $3.11\pm0.10$   &$1.07\pm0.17$                   & $\Omega_c(3119)$\\ \hline \hline
\end{tabular}
\end{center}
\caption{  The  masses $M$,  pole residues $\lambda$ and possible assignments of the charmed baryon states, where the $j_l$ denotes the total angular momentum of the light degree of freedom.}
\end{table}

\begin{figure}
 \centering
 \includegraphics[totalheight=5cm,width=7cm]{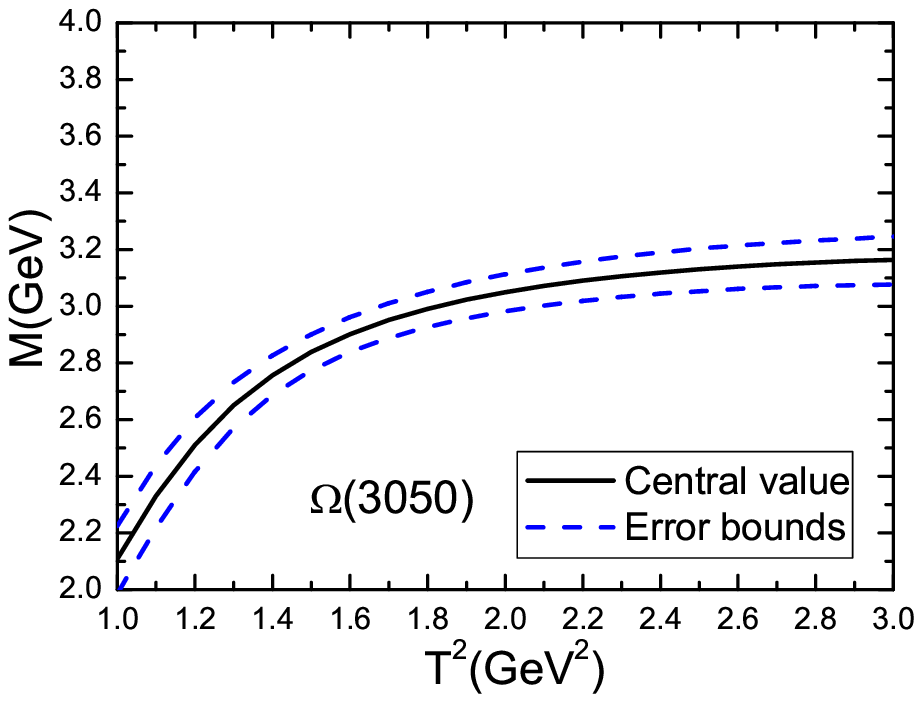}
 \includegraphics[totalheight=5cm,width=7cm]{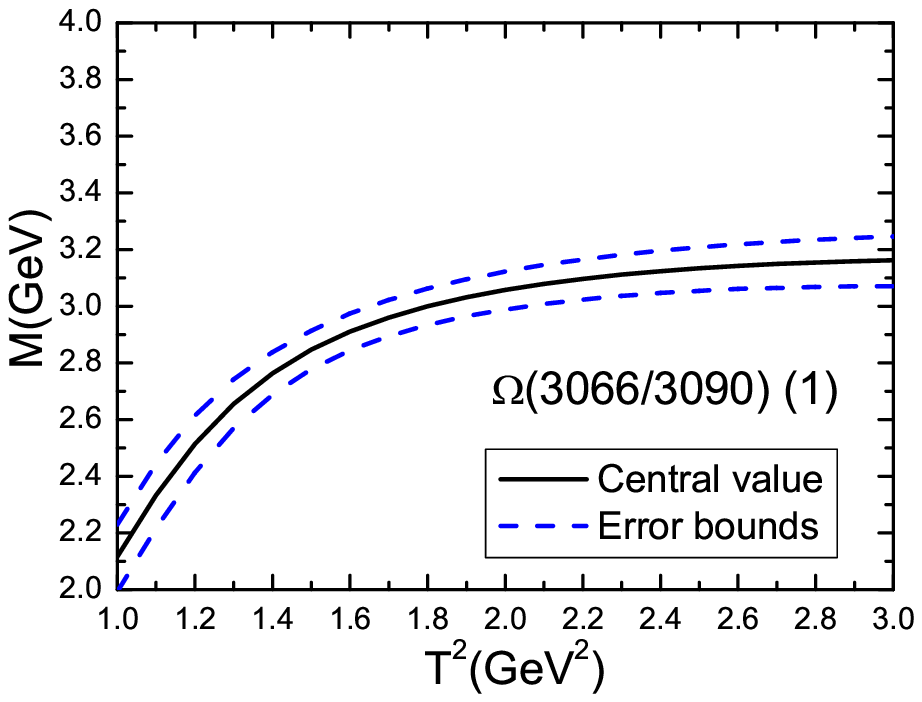}
 \includegraphics[totalheight=5cm,width=7cm]{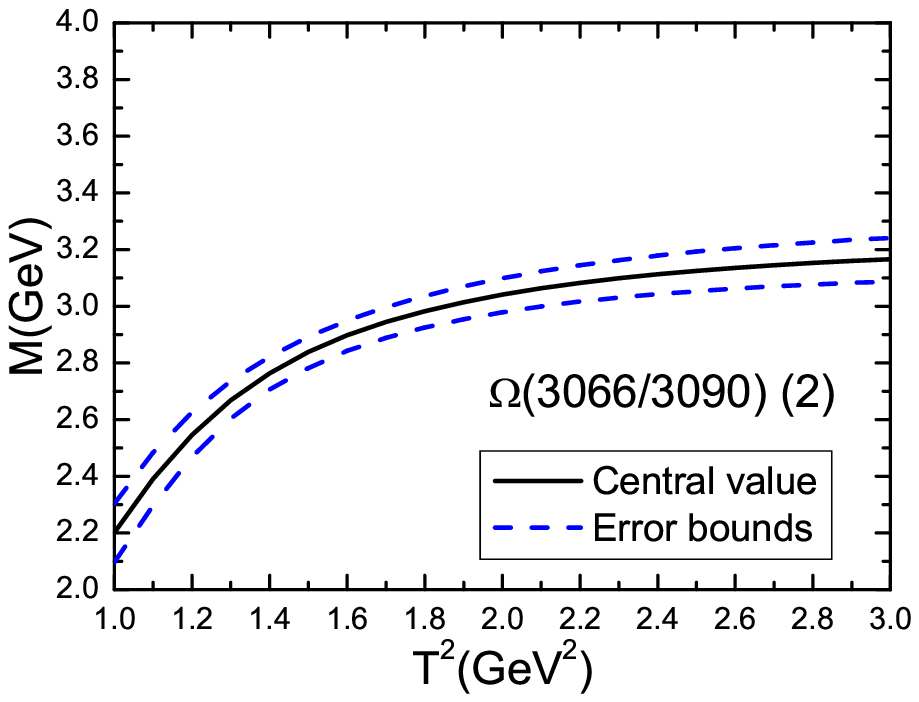}
 \includegraphics[totalheight=5cm,width=7cm]{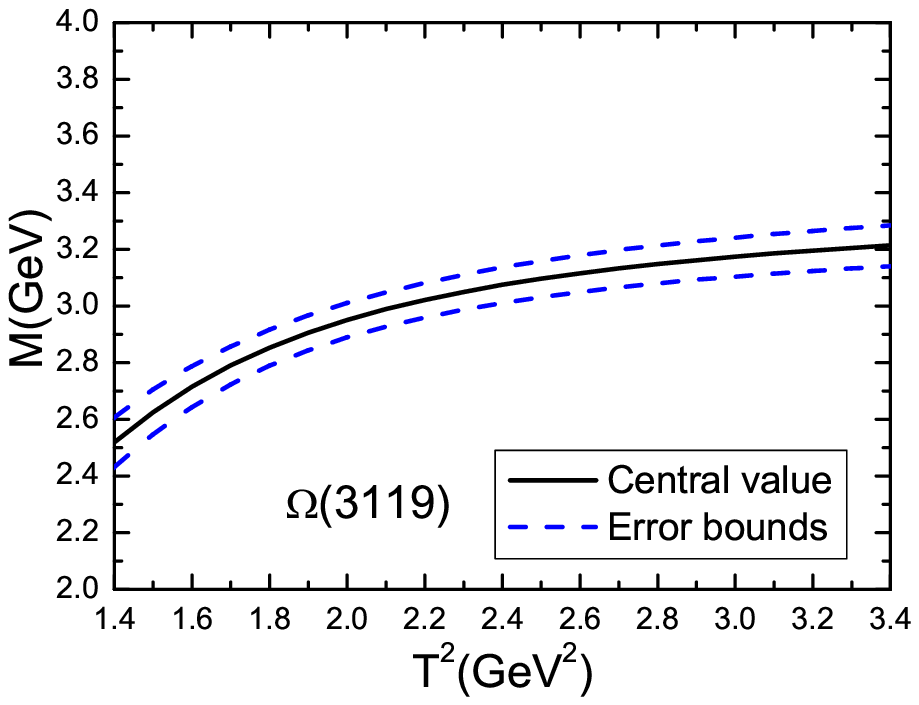}
        \caption{ The masses  of the charmed baryon states  with variations of the Borel parameters $T^2$, where the $(1)$ and $(2)$ correspond to the currents $J^1_\mu$ and $J^2_\mu$, respectively.  }
\end{figure}

\begin{figure}
 \centering
 \includegraphics[totalheight=5cm,width=7cm]{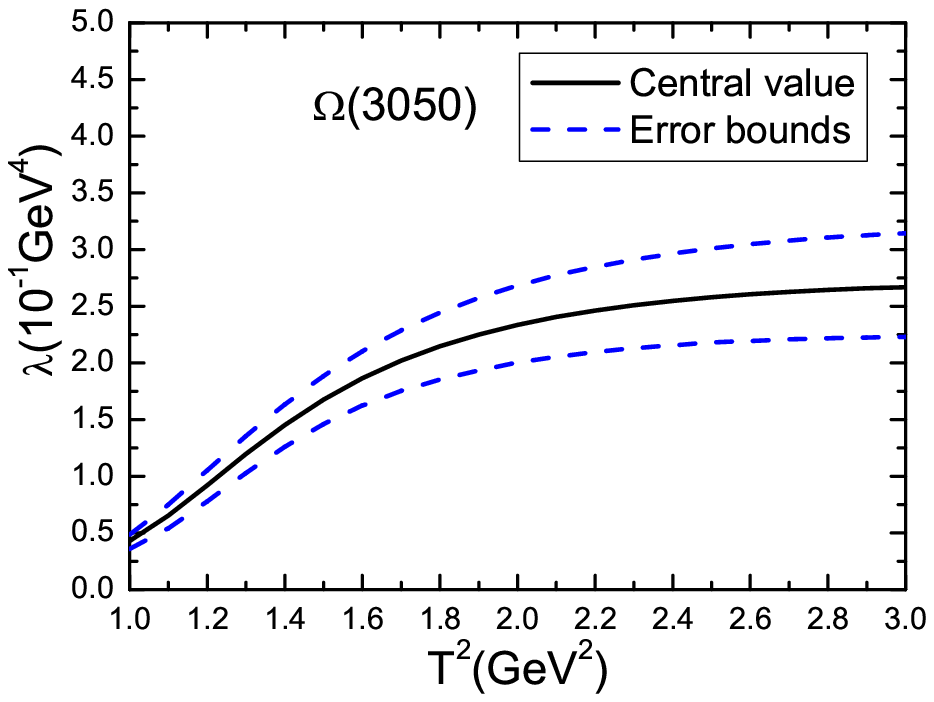}
 \includegraphics[totalheight=5cm,width=7cm]{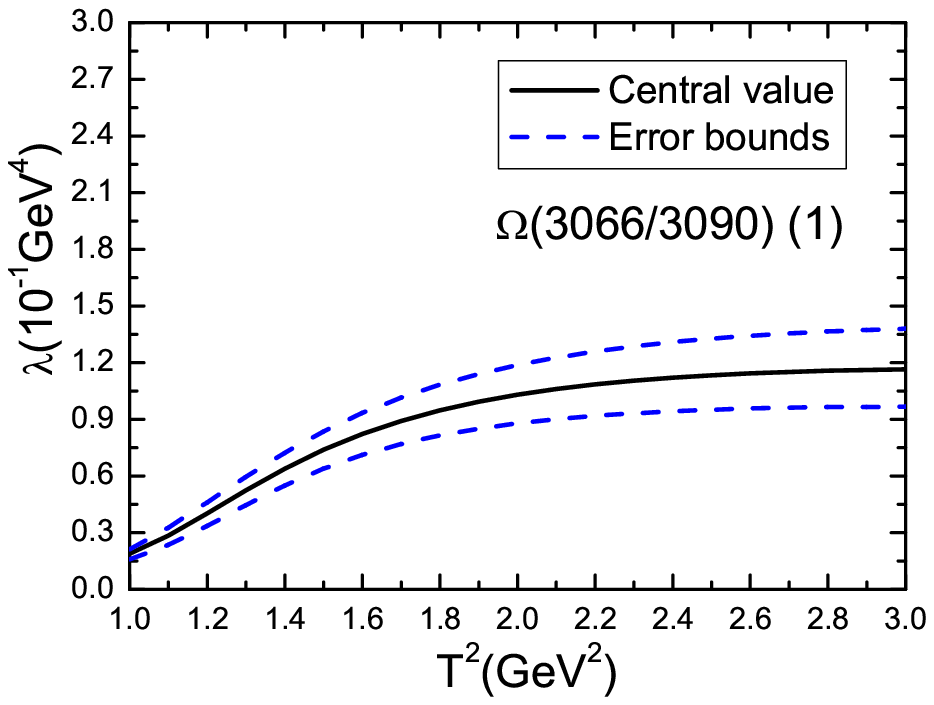}
 \includegraphics[totalheight=5cm,width=7cm]{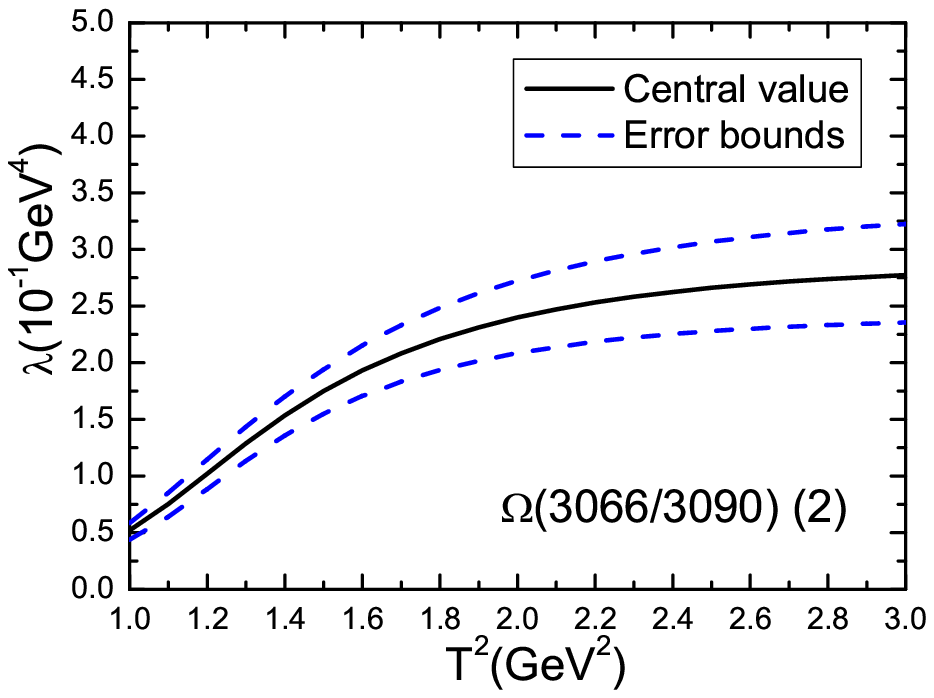}
 \includegraphics[totalheight=5cm,width=7cm]{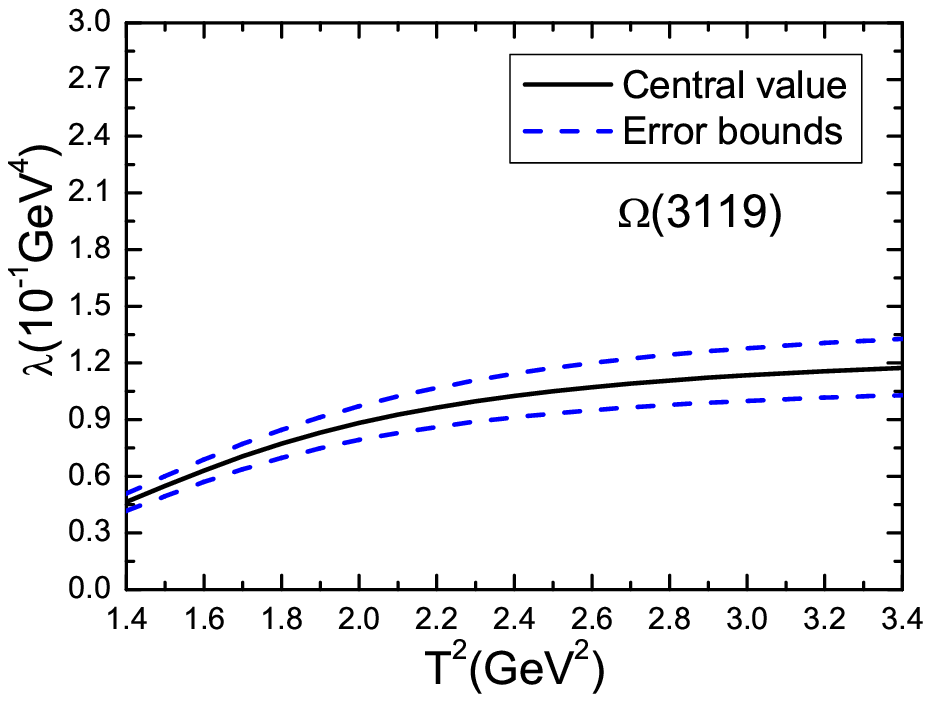}
        \caption{ The pole residues  of the charmed baryon states  with variations of the Borel parameters $T^2$, where the $(1)$ and $(2)$ correspond to the currents $J^1_\mu$ and $J^2_\mu$, respectively.   }
\end{figure}

We take into account all uncertainties  of the relevant   parameters,
and obtain the values of the masses and pole residues of
 the  $\Omega_c^0$ baryon states, which are shown in Figs.1-2 and
Table 2. In Figs.1-2,  we plot the masses and pole residues with variations
of the Borel parameters at much larger intervals   than the  Borel windows shown in Table 1. In the Borel windows, the uncertainties originate from the Borel parameters in the Borel windows are very small, $\delta M_{\Omega_c}/M_{\Omega_c}=(1.2-1.6)\%$, the criterion $\bf 3$ is also satisfied. The three  criteria are all satisfied, we expect to make reliable predictions. In Figs.1-2 and Table 2, we also present the possible assignments of the $\Omega_c^0$ states according to the masses.

In Ref.\cite{WangNegativeP}, we choose the currents without introducing relative P-wave to study the negative parity heavy and doubly-heavy   baryon states, and obtain the predictions $M=2.98\pm0.16 \,\rm{GeV}$ for the $\Omega_c^0$ states with $J^P={\frac{1}{2}}^-$, ${\frac{3}{2}}^-$, where the diquark constituent
$\varepsilon^{ijk}s^T_jC\gamma_\mu s_k$ is taken to construct the currents. Multiplying $i \gamma_{5}$ to the baryon currents  changes their parity, we can choose currents  without introducing relative P-wave to study the P-wave baryon states.
The current $\varepsilon^{ijk}s^T_jC\gamma_\mu s_k \gamma^\mu c_k$ couples potentially to the $\Omega_c^0$ state with $J^P={\frac{1}{2}}^-$ \cite{WangNegativeP}, the  mass of the $\Omega_c(3000)$ is in excellent agreement with the prediction $M=2.98\pm0.16 \,\rm{GeV}$ \cite{WangNegativeP} or the prediction $M=2.990\pm0.129\,\rm{GeV}$ based on a more general interpolating current with additional parameter \cite{Agaev17}, the  $\Omega_c(3000)$ can be assigned to the P-wave charmed baryon state with $J^{P}={\frac{1}{2}}^-$,  where two $s$ quarks are in relative S-wave.  In Table 3, we present some predictions for the masses of the P-wave $\Omega_c^0$ baryon states from the full QCD sum rules \cite{WangNegativeP,Agaev17,Azizi15} and potential quark models \cite{Shah16,Roberts07,Valcarce08,Ebert11}. We cannot identify a baryon state unambiguously with the mass alone, it is necessary to study the decay widths of those P-wave baryon states with the QCD sum rules. In Ref.\cite{Agaev17}, Agaev,  Azizi and  Sundu study the masses and widths of the 1P ${\frac{1}{2}}^-$, ${\frac{3}{2}}^-$ and 2S ${\frac{1}{2}}^+$, ${\frac{3}{2}}^+$ $\Omega_c^0$ baryon states with the full QCD sum rules, and assign the  $\Omega_c(3000)$, $\Omega_c(3050)$ and $\Omega_c(3119)$  to be the $\Omega_c^0$ baryon states with the quantum numbers $({\rm 1P},\,\frac{1}{2}^{-})$,
 $({\rm 1P},\,\frac{3}{2}^{-})$ and $({\rm 2S},\,\frac{3}{2}^{+})$, respectively, and  assign the $\Omega_c(3066)$ or  $\Omega_c(3090)$ to be the $\Omega_c^0$ baryon state with the quantum numbers $({\rm 2S},\,\frac{1}{2}^{+})$.

\begin{table}
\begin{center}
\begin{tabular}{|c|c|c|c|c|c|c|c|c|}\hline\hline
 $J^P_{j_l}$             &This Work       &\cite{WangNegativeP} &\cite{Agaev17}  &\cite{Azizi15} &\cite{Shah16} &\cite{Roberts07} &\cite{Valcarce08} &\cite{Ebert11}        \\ \hline
${\frac{1}{2}}^{-}_{0}$  &$3.05\pm0.11$   &                     &                &               &3.011          &2.977            &3.035              &3.055  \\ \hline

${\frac{1}{2}}^{-}_{1}$  &                &$2.98\pm0.16$        &$2.990\pm0.129$ &               &3.028          &2.990            &3.125              &2.966  \\ \hline

${\frac{3}{2}}^{-}_{1}$  &$3.06\pm0.11$   &$2.98\pm0.16$        &$3.056\pm0.103$ &$3.08\pm 0.12$ &2.976          &2.986            &                   &3.054  \\ \hline

${\frac{3}{2}}^{-}_{2}$  &$3.06\pm0.10$   &                     &                &               &2.993          &2.994            &                   &3.029  \\ \hline

${\frac{5}{2}}^{-}_{2}$  &$3.11\pm0.10$   &                     &                &               &2.947          &3.014            &                   &3.051  \\ \hline
\end{tabular}
\end{center}
\caption{ The masses of the P-wave $\Omega_c$ baryon states, where the unit is GeV, the $j_l$ denotes the total angular momentum of the light degree of freedom. We neglect the mixing effects of the ${\frac{1}{2}}^{-}_{0}-{\frac{1}{2}}^{-}_{1}$ and ${\frac{3}{2}}^{-}_{1}-{\frac{3}{2}}^{-}_{2}$ in the potential quark models for simplicity. }
\end{table}

\section{Conclusion}
In this article, we assign the $\Omega_c(3000)$, $\Omega_c(3050)$, $\Omega_c(3066)$, $\Omega_c(3090)$ and $\Omega_c(3119)$  to be the  P-wave charmed baryon states with $J^P={\frac{1}{2}}^-$, ${\frac{1}{2}}^-$, ${\frac{3}{2}}^-$, ${\frac{3}{2}}^-$ and ${\frac{5}{2}}^-$, respectively, and study their masses and pole residues  with the QCD sum rules in details by introducing an explicit relative P-wave between the two constituents  of the light diquarks.  The predictions support assigning the $\Omega_c(3050)$, $\Omega_c(3066)$, $\Omega_c(3090)$ and $\Omega_c(3119)$  to be the  P-wave baryon states with $J^P={\frac{1}{2}}^-$, ${\frac{3}{2}}^-$, ${\frac{3}{2}}^-$ and ${\frac{5}{2}}^-$, respectively, where the two constituents  of the light diquark are in relative P-wave; while the  $\Omega_c(3000)$ can be assigned to the P-wave charmed baryon state with $J^{P}={\frac{1}{2}}^-$, where the two constituents  of the light diquark are in relative S-wave.

\section*{Acknowledgements}
This  work is supported by National Natural Science Foundation, Grant Number 11375063.

\end{document}